\documentclass[aps,prx,reprint,superscriptaddress]{revtex4-1}
\usepackage{amsmath,amsthm,amssymb,bm,comment}
\usepackage{graphicx}

\graphicspath{{./fig/}}

\begin{document}


\title{Disordered quantum spin state in the stripe lattice system\\consisting of triangular and square tilings}


\author{Y. Saito}
\email[]{sai10@phys.sci.hokudai.ac.jp}
\affiliation{Department of Physics, Hokkaido University, Sapporo 060-0810, Japan}
\affiliation{1. Physikalisches Institut, Universit\"{a}t Stuttgart, Pfaffenwaldring 57, 70569 Stuttgart}

\author{H. Nakamura}
\affiliation{Department of Physics, Hokkaido University, Sapporo 060-0810, Japan}

\author{M. Sawada}
\affiliation{Department of Physics, Hokkaido University, Sapporo 060-0810, Japan}

\author{T. Yamazaki}
\affiliation{Department of Physics, Hokkaido University, Sapporo 060-0810, Japan}

\author{S. Fukuoka}
\affiliation{Department of Physics, Hokkaido University, Sapporo 060-0810, Japan}

\author{N. Matsunaga}
\affiliation{Department of Physics, Hokkaido University, Sapporo 060-0810, Japan}

\author{K. Nomura}
\affiliation{Department of Physics, Hokkaido University, Sapporo 060-0810, Japan}

\author{M. Dressel}
\affiliation{1. Physikalisches Institut, Universit\"{a}t Stuttgart, Pfaffenwaldring 57, 70569 Stuttgart}

\author{A. Kawamoto}
\affiliation{Department of Physics, Hokkaido University, Sapporo 060-0810, Japan}

\begin{abstract}
Quantum fluctuations originating phase competition or geometrical frustration of spins lead to novel states such as a quantum critical point and a quantum spin liquid where the strong quantum fluctuations suppress any ordered states even at 0 K. Utilizing site-selective NMR for a quasi-two dimensional organic conductor $\lambda$-(STF)$_2$GaCl$_4$, we investigate the non-magnetic insulating phase of the stripe lattice system consisting of triangular and square tilings. We found development of AF spin fluctuations with decreasing temperature. Regardless of large enhancement of spin-lattice relaxation rate $1/T_1$ owing to critical slowing down below 10 K, no long-range magnetic ordering was observed down to 1.63 K two orders of magnitude less than the exchange interaction $J/k_{\rm B} \simeq$ 194 K. Moreover, $1/T_1$ saturated below 3.5 K. These results are in stark contrast to observed behaviors so far in other non-magnetic ground states discussed in terms of spin liquids, demonstrating realization of an exotic quantum state accompanying quantum criticality.
\end{abstract}


\maketitle


\section{INTRODUCTION}
Strong quantum fluctuations give birth to novel ground states and low-lying energy excitation, so called quantum critical phenomena realized near a quantum critical point (QCP) or in quantum spin liquids \cite{Sachdev2011,Haravifard2015,Shibauchi2014,Balents2010,Knolle2019,Takagi2019}. Strong spin fluctuations near a QCP are regarded as a key to clarify the unconventional superconducting (SC) pairing mechanism \cite{Stewart2017,Bennemann}. Furthermore, a SC phase adjacent to a spin liquid phase may have an exotic SC mechanism like Amperean or spin-triplet pairing \cite{Lee2007, Galitski2007}. Therefore, investigating quantum critical phenomena is one of the intriguing topics in solid state physics unveiling not only the quantum state itself but also the relation to the unconventional SC pairing mechanism. However, such a quantum state is usually masked by another phase or difficult to access because of precise tuning parameters, turning out to be challenging and cumbersome to investigate properties of the quantum states.

\begin{figure*}[tbp]
  \includegraphics[width=17.2cm]{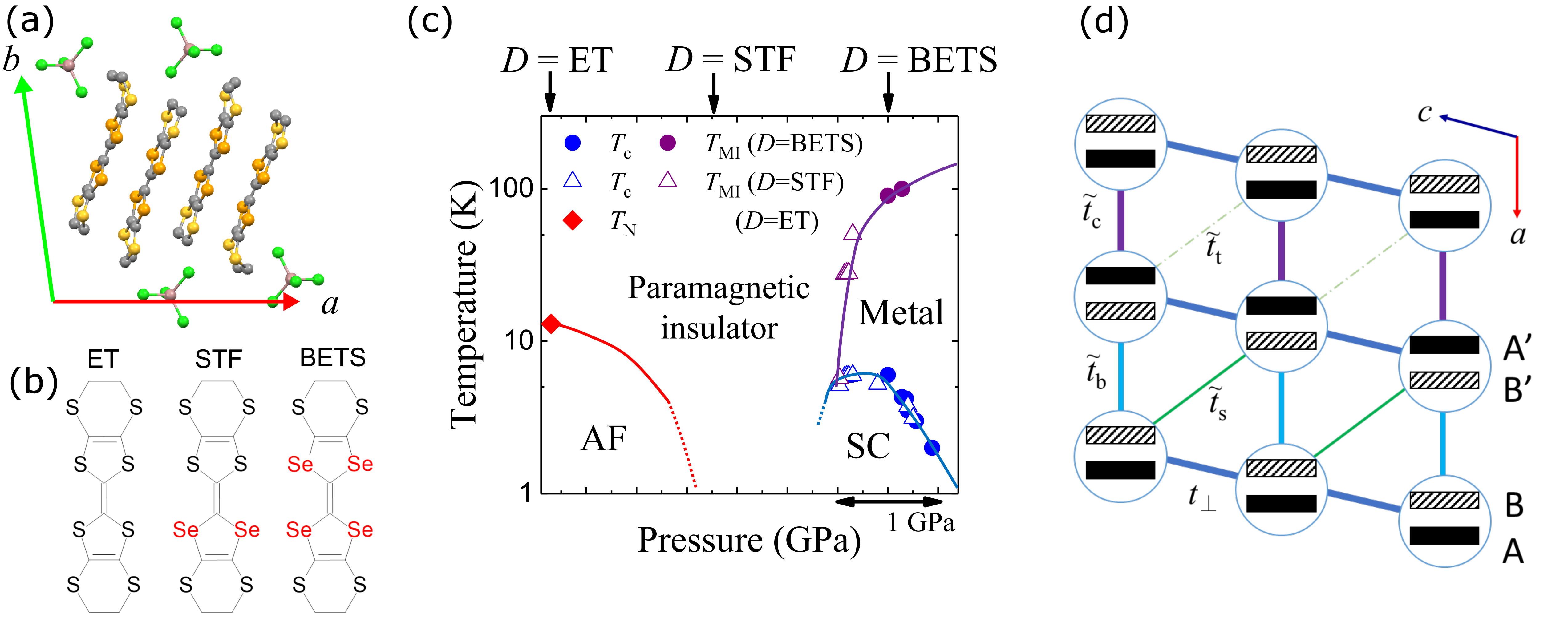}
  \caption{(a) Crystal structure of $\lambda$-$D_2$GaCl$_4$. $\lambda$-type salts consist of donor molecules $D$ with the tetrahedral anion GaCl$_4$ and form charge-transfer salts with two-dimensional layered structures. (b) Donor molecular structure of ET (BEDT-TTF) [bis(ethylenedithio)tetrathiafulvalene], STF (BEDT-STF) [unsymmetrical-bis(ethylenedithio)diselenadithiafulvalene], and BETS (BEDT-TSF) [bis(ethylenedithio)tetraselenafulvalene]. (c) Pressure-temperature phase diagram of $\lambda$-$D_2$GaCl$_4$. $T_{\mathrm c}$, $T_{\mathrm {MI}}$, and $T_{\mathrm N}$ represent the superconducting transition temperature, the metal--insulator transition temperature, and the N\'eel temperature, respectively. The data are taken from Refs.\cite{Tanaka1999, Mori2001, Minamidate2015,Saito2018}. (d) Donor molecular arrangement with transfer integrals among dimers in $\lambda$-type salts. Donor molecules A (B) and A' (B') are associated with inversion symmetry. Circles represent dimers. Because $\tilde{t}_t$ is much smaller than other $t$, the effective exchange interaction network consists of the stripe lattice of triangular (lower parts) and square (upper parts) tilings. See for details in Sec. IV B.}\label{crystal}
\end{figure*}

Quasi-two dimensional organic conductor $\lambda$-$D_2$GaCl$_4$ has attracted much attention because of rich electronic properties such as frustrated spin magnetism \cite{Minamidate2015} and unconventional superconductivity \cite{Imajo2019} including Fulde-Ferrell-Larkin-Ovchinnikov state under high magnetic field \cite{Coniglio2011,Uji2015}. The substitution of a donor molecule $D$ from ET to STF and BETS gives rise to the chemical pressure effect and tunes the electronic state as shown in Fig. \ref{crystal}(a-c) \cite{Mori2001, Minamidate2015, Saito2018}.
The magnetic susceptibility of $\lambda$-(STF)$_2$GaCl$_4$, which is situated between the AF insulating phase and the SC phase, follows the spin-1/2 AF Heisenberg model on a triangular lattice and no magnetic ordering down to 2 K two orders of magnitude less than the exchange interaction \cite{Minamidate2015}, suggesting possibility of a pressure-induced spin liquid. Distinct features of $\lambda$ salts are their lattice structure is combination of triangular and square tilings \cite{Tanaka1999, Seo2004} as shown in Fig. \ref{crystal}(d), and a non-magnetic state exists between the AF phase and the SC phase in contrast to well-known organic conductor $\kappa$-type salts of which SC phase is adjacent to the AF or the spin liquid candidate phase \cite{And2004,Kawamoto2006,Shimizu2016}. Moreover, it is suggested that $\lambda$ salts are close to the phase boundary between AF and spin gap phases \cite{Seo1997}. Therefore, the magnetic properties of the stripe lattice system is of significant research interest for the relation to quantum states.

In this paper, we investigate no-magnetic phase realized in the stripe lattice system consisting of triangular and square tilings using $\lambda$-(STF)$_2$GaCl$_4$. We performed powerful site-selective $^{13}$C NMR, a microscopic probe that measures the static and dynamic magnetic properties, where the selected $^{13}$C sites are the best nuclei to investigate the electronic state \cite{Hirose2012}. We found that observed spin dynamics is distinct from spin liquids, demonstrating an exotic quantum state near a competing magnetic phase in the stripe lattice system.

\section{EXPERIMENTAL}
Single crystals of $\lambda$-(STF)$_2$GaCl$_4$ were prepared from the electrochemical oxidation of STF in chlorobenzene in the presence of tetra-$n$-butylammonium (TBA) GaCl$_4$ \cite{Naito1997}. STF was synthesized according to a literature \cite{Naito1997}. Importantly, S side of the central C=C bond in STF molecules was enriched with $^{13}$C isotope \cite{Hirose2012} to avoid NMR spectrum splitting due to dipolar coupling \cite{Pake1948}. NMR experiments were performed for a single crystal in a magnetic field of 6 T for spectrum measurements and 7 T for spin-lattice relaxation rate $1/T_1$ and spin-spin relaxation rate $1/T_2$. The orientation of the magnetic field $\theta$ was defined by angle from the magnetic field parallel to the [110] direction which is nearly parallel to a long axis of a STF molecule to the $-b'\: (\equiv -c\times a)$ axis. NMR spectra were obtained by the fast Fourier transformation of the spin echo signal with a $\pi/2{\mathrm -}\pi$ pulse sequence. The NMR shifts are given in ppm which is relative to tetramethylsilane (TMS). $1/T_1$ was measured using the conventional saturation-recovery method. The linewidths were evaluated by fitting peaks to the Lorentz function and $1/T_2$ was defined as the rate corresponding to Lorentz decay. DC resistivity was measured along the $c$ axis from room temperature down to 79 K by the standard four-point probe technique.

\section{RESULTS}
\subsection{NMR spcetra, linewidth and spin susceptibility}
\begin{figure}[bp]
  \includegraphics[width=8.6cm]{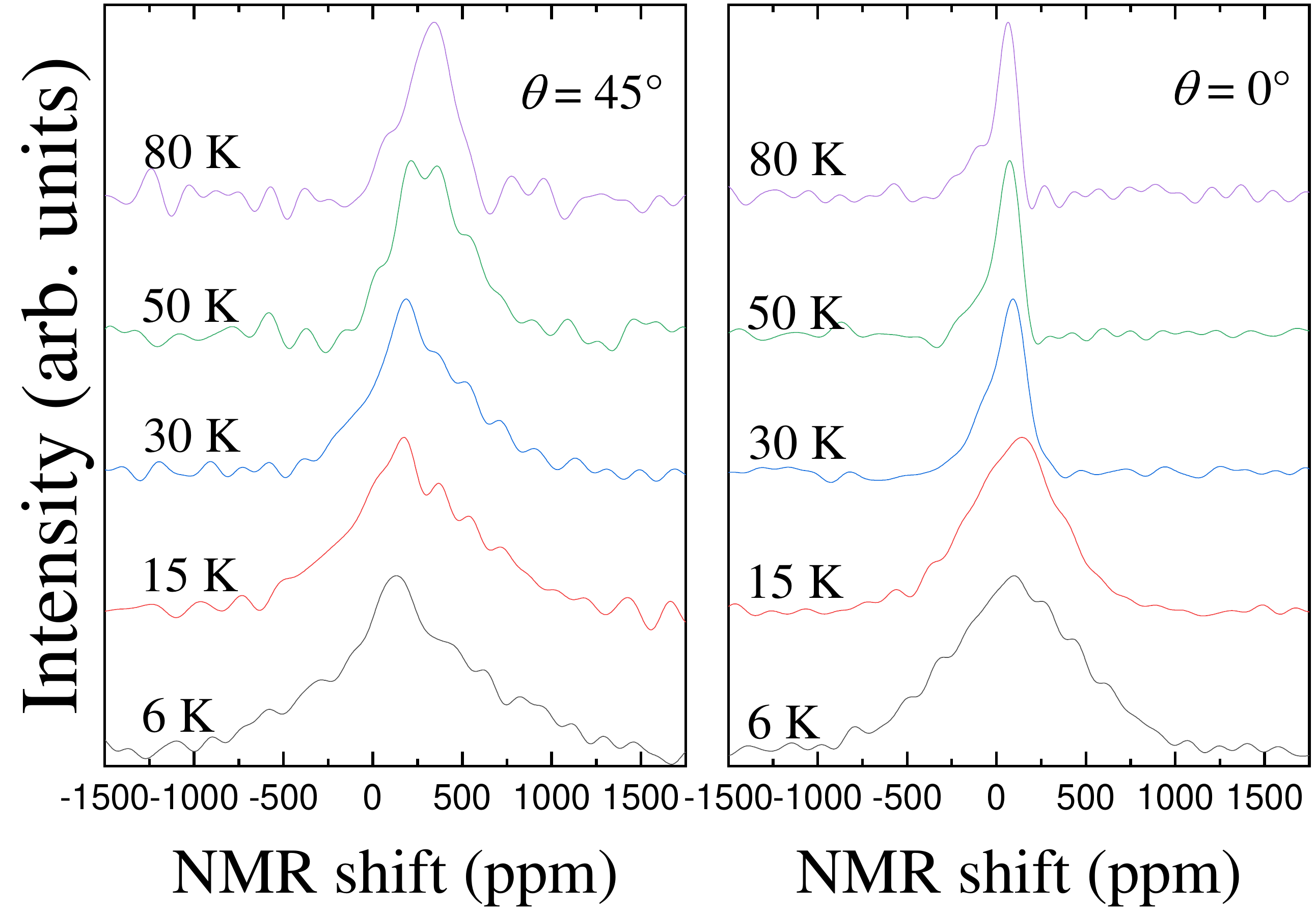}
  \caption{NMR spectrum at several temperatures of $\lambda$-(STF)$_2$GaCl$_4$ at $\mu_0 H$ = 6 T for $\theta=45^{\circ}$ (left panel) and 0$^{\circ}$ (right panel).}\label{spec}
\end{figure}
$\lambda$ salts have two crystallographically independent donor molecules A (A') and B (B'), where A (B) and A' (B') molecules are associated with inversion symmetry, because the crystal structure belongs to $P\overline{1}$. Since each molecule has two central carbon sites, so called inner and outer sites \cite{Saito2015}, four resonance lines are expected. Figure \ref{spec} shows NMR spectra at several temperatures for magnetic field 6 T applied to $\theta=0^{\circ}$ (almost parallel to a long axis of a STF molecule) and 44.7$^{\circ}$. A broad peak with unsymmetrical line shape was observed at both angles, suggesting that expected four peaks were merged to one broad peak. The NMR shift for $H$ parallel to 45$^{\circ}$ decreased with decreasing temperature, whereas the NMR shift for $H$ parallel to 0$^{\circ}$ increased, indicating each hyperfine coupling constant has the opposite sign. The hyperfine coupling constants were estimated as $A_{45^{\circ}} = 2.1$ kOe/$\mu_\text{B}$ and $A_{0^{\circ}}= 0.46$ kOe/$\mu_\text{B}$ by angle dependence of NMR shift (see Appendix A). Since there was no clear peak splitting down to the lowest temperature, there is no indication of emergence of internal field due to long-range magnetic ordering consistent with magnetic susceptibility measurements \cite{Minamidate2015}.

\begin{figure}[tbp]
  \includegraphics[width=8.6cm]{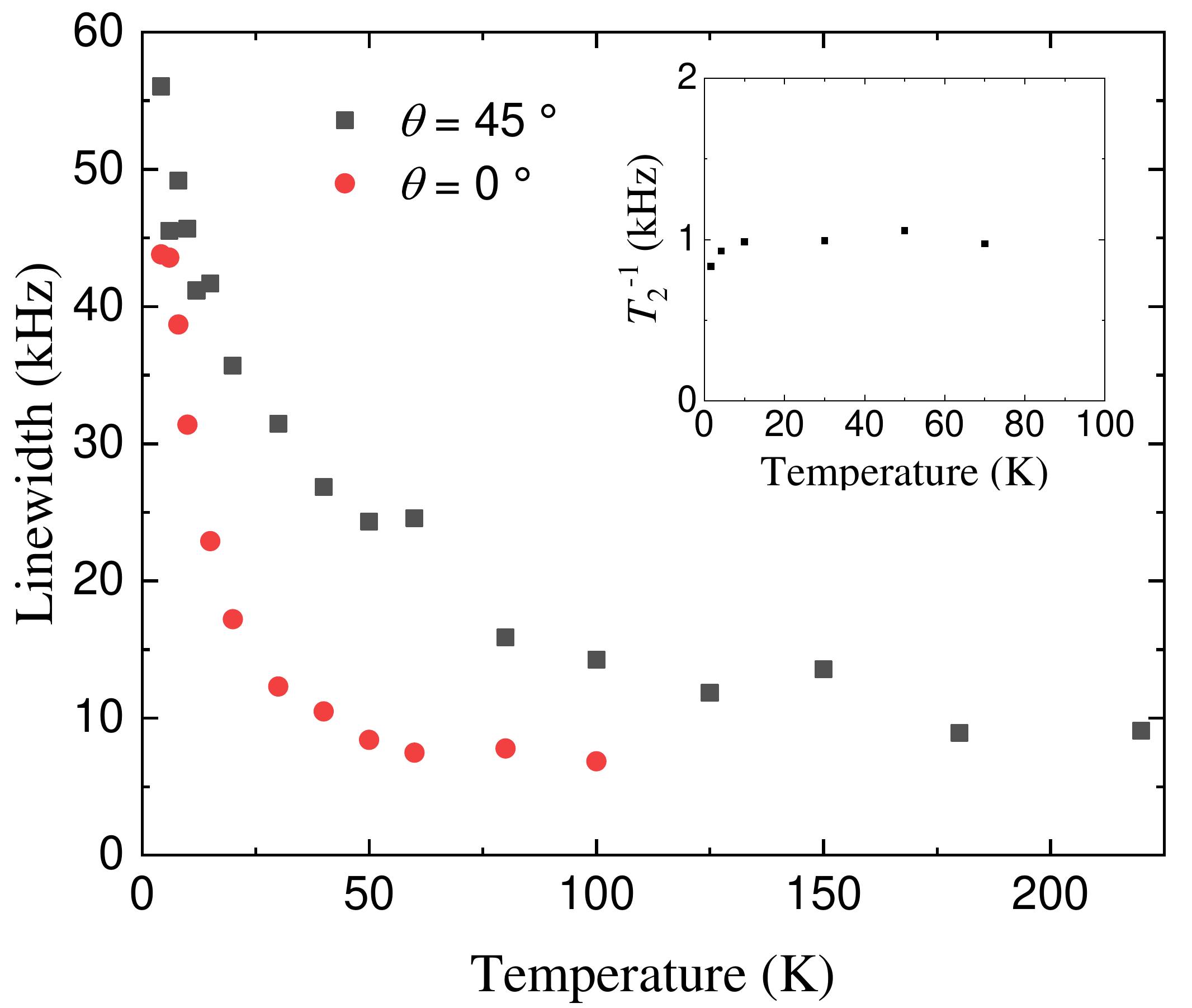}
  \caption{Temperature dependence of FWHM of $\lambda$-(STF)$_2$GaCl$_4$ at $\mu_0 H$ = 6 T for $\theta=45^{\circ}$ and 0$^{\circ}$. The inset shows $T_2^{-1}$ at $\mu_0 H$ = 7 T for $\theta=-12.8^{\circ}$.}\label{FWHM}
\end{figure}

Figure \ref{FWHM} shows temperature dependence of the full width at half maximum (FWHM) of the NMR lines. As lowering temperature, The linewidth gradually broadened and the broadening was clearly seen for both angles below 40 K. Further broadening was observed Below 20 K. For Fourier transform NMR, the linewidth $\Delta \omega$ of the spectrum is generally described as
\begin{equation}
\Delta \omega = \frac{2 \pi}{T_2}+\gamma_{\mathrm I} \Delta H \label{T2FWHM}
\end{equation}
where $\gamma_{\mathrm I}$ is the nuclear gyromagnetic ratio and $\Delta H$ is the inhomogeneity of the local magnetic field at the corresponding nuclei. The first term is inhomogeneous width caused by slow magnetic fluctuations, and the second term is a static inhomogeneous width caused by the inhomogeneity of the external and local magnetic field. $1/T_2$ can detect the slow magnetic fluctuations of $\simeq$ 10 kHz. Since $1/T_2$ remains constant at all temperatures as shown in the inset of Fig. \ref{FWHM}, indicating that the broadening originates not from dynamics but from inhomogeneous broadening.

\begin{figure}[tbp]
  \includegraphics[width=8.6cm]{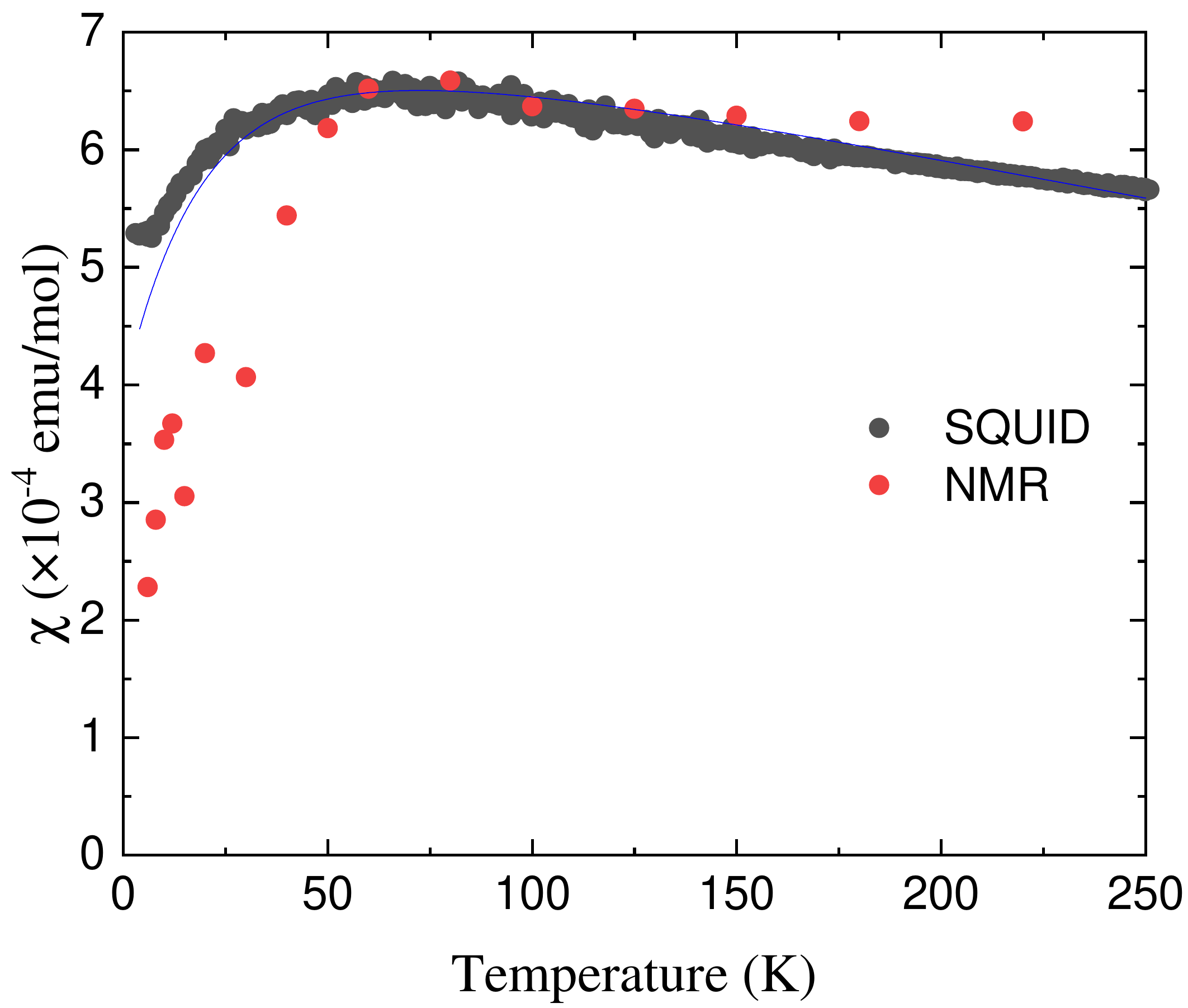}
  \caption{Temperature dependence of magnetic susceptibility obtained from SQUID and NMR for $\lambda$-(STF)$_2$GaCl$_4$. The SQUID data are taken from Ref.\cite{Minamidate2015} and the core diamagnetic contribution is re-estimated. The solid curve is the fit of the $S=1/2$ AF Heisenberg model on a triangular lattice with $J=194$ K.} \label{static}
\end{figure}

Figure \ref{static} shows temperature dependence of spin susceptibility $\chi_{\mathrm {SQUID}}$ and $\chi_{\mathrm{NMR}}$ obtained by SQUID \cite{Minamidate2015} and NMR measurements for $\theta =44.7^{\circ}$. Here, the core diamagnetic contribution of the SQUID measurement is re-estimated to be $-4.82 \times10^{-4}$ emu/mol f.u. and already subtracted. The solid curve shows the fit of the high-temperature series expansion of a spin $S=1/2$ two-dimensional AF Heisenberg model on the triangular lattice using [7/7] Pad\'{e} approximants \cite{Tamura2002}, where the exchange interaction is 194 K, the effective moment is 0.97 $\mu_\text{B}$, and $g$ is 2.0104 obtained by electron spin resonance. Both susceptibility $\chi_{\mathrm s}$ showed similar behavior at high temperature, however, $\chi_{\mathrm {NMR}}$ deviated from $\chi_{\mathrm {SQUID}}$ at low temperature. The similar trend of $\chi_{\mathrm {NMR}}$ was also confirmed by the results of $\theta =0^{\circ}$ (not shown).

\subsection{Nuclear spin-lattice relaxation rate $1/T_1$}
\begin{figure}[tbp]
  \includegraphics[width=8.6cm]{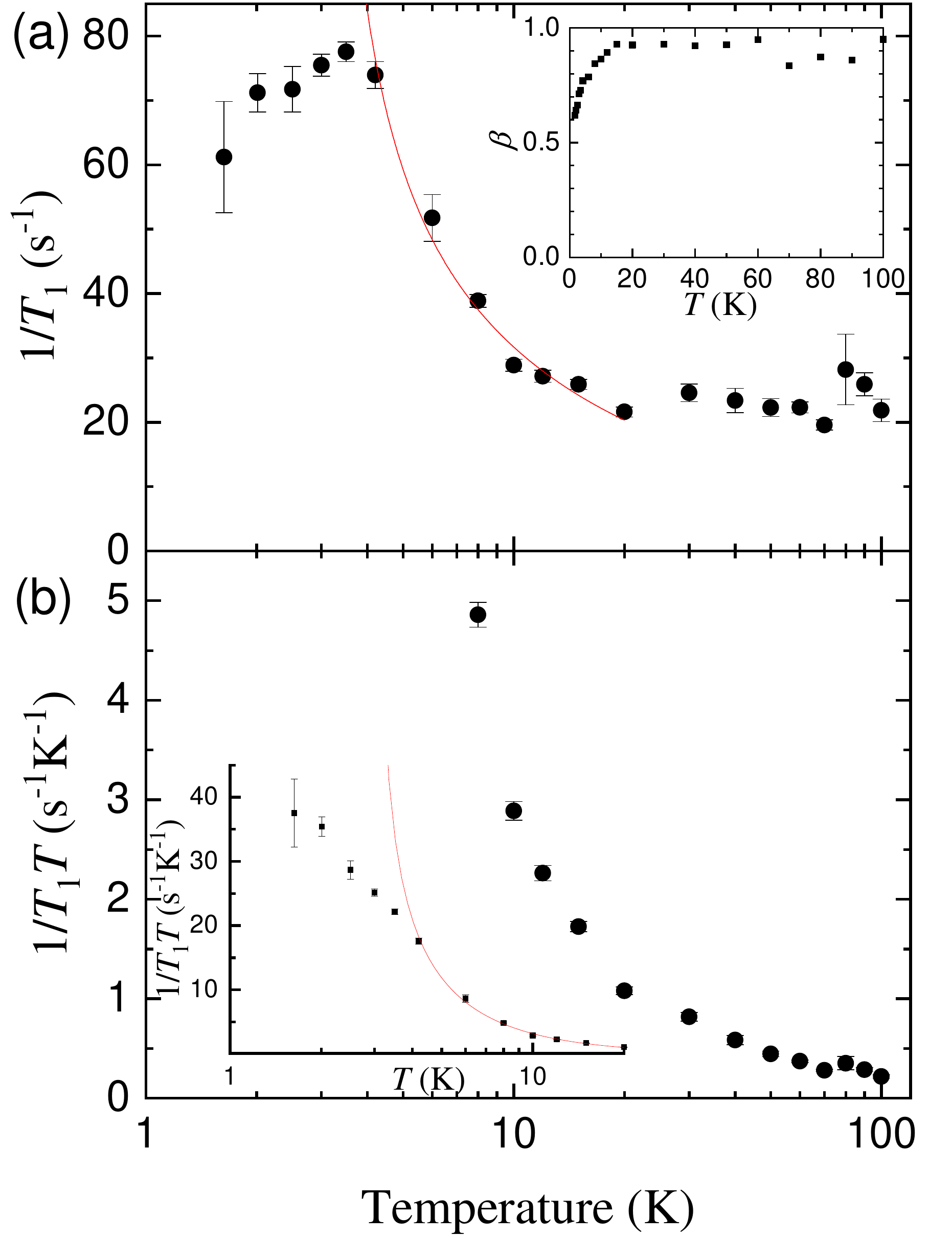}
  \caption{Temperature dependence of (a) $1/T_1$ and (b) $1/T_1T$ of $\lambda$-(STF)$_2$GaCl$_4$ at $\mu_0 H$ = 7 T for $\theta=-12.8^{\circ}$. The inset (a) shows the exponent $\beta$ in the stretched exponential fit. $1/T_1T$ below 20 K are enlarged in the inset (b). The solid curves are $1/T_1\propto (T-T_\text{c})^{-1/2}$.}\label{T1T}
\end{figure}

$1/T_1$ detects low-energy excitation of electronic states and $1/T_1T$ is useful to discuss dynamic property of magnetism because $1/T_1T$ is proportional to the imaginary part of the dynamic susceptibility $\sum_{\bm{q}} \chi'' (\bm{q},\: w)/w$ at wave vector $\bm{q}$ and NMR frequency $w$. Figure \ref{T1T} shows (a) $1/T_1$ and (b) $1/T_1T$ as a function of temperature. $T_1$ was evaluated from the single-stretched-exponential recovery, $1-M(t)/M_0={\mathrm {exp}} [1-(t/T_1)^\beta]$ where $M(t)$ is the nuclear magnetization, and $M_0$ is a saturated value of $M(t)$. $\beta$ represents homogeneity of $1/T_1$ and deviation from 1 indicates inhomogeneity of $1/T_1$. $1/T_1$ is almost constant from 100 to 20 K, as expected in a localized spin system well above the N\'{e}el temperature, where $\beta$ is almost 1. Slight deviation of $\beta$ from 1 is due to multiple components of merged NMR spectra. Below 10 K, $1/T_1$ shows anomalous increase, indicating critical slowing down toward a magnetic transition. $\beta$ started to decrease below 10 K, which suggests distribution of $1/T_1$. Temperature dependence of $1/T_1$ and $1/T_1T$ above 4.2 K followed the equation expected above critical temperature $T_{\text c}$ in an AF localized electron system, $1/T_1\propto (T-T_\text{c})^{-1/2}$, where we used $T_{\text c}=3$ K. 

Below 3.5 K, $1/T_1$ was saturated and decreased slightly at the lowest temperature. $1/T_1T$ increased with decreasing temperature which is characteristic of development of AF spin fluctuations. No divergence peak in both $1/T_1$ and $1/T_1T$ indicates absence of long-range magnetic ordering which is consistent with no peak splitting of NMR spectra in contrast to $\lambda$-(ET)$_2$GaCl$_4$. 

\section{DISCUSSION}
\subsection{Dimer--Mott insulator model on $\lambda$-type salts}
Here, we consider the electronic structure of $\lambda$ salts to clarify the origin of AF spin fluctuations. Band calculations show the upper two and lower two bands with the energy gap due to the dimerization in $\lambda$ salts \cite{Tanaka1999,Mori2001,Aizawa2018}. Because GaCl$_4^{-1}$ is a monovalent anion and the formal charge of a donor molecule is +0.5, the lower two bands are filled and the upper two bands are half-filled. Hence, the electronic system is regarded as the half-filled system, so-called the dimer-Mott system as established in $\lambda$-(ET)$_2$GaCl$_4$. On a dimer-Mott system, the electronic system shows insulating behavior due to on-site Coulomb repulsion $U$ and AF fluctuations from the exchange interaction $J$ is expected. We evaluates $J$ from $1/T_1$ using Moriya's expression \cite{Relaxation1956} in the same manner as $\lambda$-(ET)$_2$GaCl$_4$ \cite{Saito2018}. From $1/T_1=24$ s$^{-1}$ which is constant in the high-temperature region, $J/k_{\mathrm B}$ was estimated to be 215 K for the square lattice and 176 K for the triangular lattice, order of which is in good agreement with $J/k_{\mathrm B}$ = 194 K from static magnetic susceptibility fit. The development of AF spin fluctuations is consistent with the dimer-Mott insulating picture which localized spins on dimers fluctuating with $J$. The qualitative similarity to $1/T_1$ above $T_{\mathrm N}=13$ K of $\lambda$-(ET)$_2$GaCl$_4$ indicates the origin of AF spin fluctuations is the same for both salts.

\subsection{Absence of magnetic ordering}
It is remarkable that long-range magnetic ordering is not observed at $T < 0.01J$ even though $1/T_1$ shows critical slowing down. Absence of a cusp structure in $1/T_1$ indicates spin freezing which is usually seen in a spin glass state is unlikely to describe the results. However, linewidth broadens gradually from high temperature. Generally, linewidth broadening is observed in spin liquid candidates. The linewidth broadening is due to static inhomogeneity since $1/T_2$ is constant at whole temperature range. Indeed, the linewidth ratio $|\nu_{45^{\circ}} /\nu_{0^{\circ}}|$ is from 2.4 to 4.1 above 30 K where natural line width of 2 kHz is already subtracted, which is close to the ratio of the hyperfine coupling constant $|A_{45^{\circ}} /A_{0^{\circ}}| =4.5$. Thus, the linewidth broadening is related to the hyperfine coupling constant, suggesting that the spin density on molecules is inhomogeneous. In other words, $\Delta K = A \Delta \chi$ (see also Appendix B). Below 20 K, further linewidth broadening occurs and deviation of $\chi_{\mathrm {NMR}}$ from $\chi_{\mathrm {SQUID}}$ is pronouced. Linewidth broadens toward opposite sign direction of the hyperfine coupling constant for both $\theta =$ 45$^{\circ}$ and 0$^{\circ}$ where the spectrum tail exceeds chemical shift of 74.6 ppm and 115.4 ppm for $\theta =$45$^{\circ}$ and 0$^{\circ}$, respectively. These results suggest staggered moments are partly induced at low temperature. This may be simply because impurities induce staggered moments. However, $\chi_{\mathrm {SQUID}}$ does not show significant Curie-like behavior at low temperature, suggesting impurity effect is not obvious.
Another explanation is Dzyaloshinsky-Moriya (DM) interaction. Similar experimental behavior is observed in an organic antiferromagnet $\kappa$-(ET)$_2$Cu[N(CN)$_2$]Cl \cite{Smith2004,Kagawa2008}. Deviation of Knight shift from $\chi_{\mathrm {SQUID}}$ and linewidth broadening occur even above $T_{\mathrm N}$ because of the field-induced moment in the short-range ordered spins by DM interaction. Recently, the importance of spin-orbit coupling (SOC) effect in organic conductors is suggested \cite{Winter2017}. SOC introduces DM vectors where SOC of Se atoms is about five times greater than that of S atoms \cite{Winter2017}. Because $\lambda$-(ET)$_2$GaCl$_4$ shows no significant line broadening above $T_{\mathrm N}$, the deviation of $\chi_{\mathrm {NMR}}$ and linewidth broadening below 20 K may be explained by the strength of SOC.

Deviation of $1/T_1$ from the relation $1/T_1\propto (T-T_\text{c})^{-1/2}$ and absence of magnetic ordering suggest some sort of quantum spin state suppresses long-range magnetic ordering. From the results that magnetic susceptibility can be fit by the two-dimensional AF Heisenberg model on a triangular lattice, possibility of spin frustration was suggested \cite{Minamidate2015}. Here we consider geometrical frustration using band calculations of the $\lambda$-BETS salt \cite{Aizawa2018, Tanaka1999}. Figure \ref{crystal}(d) shows the dimer network of $\lambda$ salts consisting of transfer integrals. The ratios of the transfer integrals obtained by first principle calculation are $|\tilde{t}_b/t_\perp|=0.96$, $|\tilde{t}_c/t_\perp|=1.01$, $|\tilde{t}_s/t_\perp|=0.6$ \cite{Aizawa2018}. Because the diagonal transfer integral $\tilde{t}_t$ in the upper lattice is about $0.1t_\perp$ which corresponds to $0.01J_\perp$ using relation $J \propto t^2/U$, upper lattices are regarded as almost square type. Here the difference between $\tilde{t}_s$ and $\tilde{t}_t$ is due to positional displacement along the long axis of the donor molecule: difference of the distance between the outer site is about 1.9 \AA. On the other hand, the extended H\"{u}ckel calculation shows $|\tilde{t}_b/t_\perp|=1.23$, $|\tilde{t}_c/t_\perp|=0.72$, $|\tilde{t}_s/t_\perp|=0.6$, and $|\tilde{t}_t/t_\perp|=0.05$ \cite{Tanaka1999}, where the square lattices are anisotropic. Both calculations indicate $\lambda$ salts consist of stripe lattices of square and triangular tilings. In principle, square lattices have no frustration without diagonal exchange interactions, though, the triangular lattices have frustration. Although $\lambda$-(ET)$_2$GaCl$_4$ shows square nature as the magnetic susceptibility is fit by the Heisenberg model on a square lattice \cite{Saito2018}, triangular nature appears by changing ET molecules to STF molecules. Therefore, the combination lattice system can have frustration. However, geometrical frustration of $\lambda$ salts is far from the ideal triangular lattice. Actually, in spin liquid candidates, for example, triangular magnets $\kappa$-(ET)$_2$Cu$_2$(CN)$_3$ and EtMe$_3$Sb[Pd(dmit)$_2$]$_2$, and a kagome magnet herbertsmithite, $1/T_1$ shows decrease without upturn at low temperature \cite{Shimizu2006,Itou2011a,Jeong2011}. Our observation, the increase of $1/T_1$ at low temperature is qualitatively different from low-temperature behavior in the frustrated spin systems. These results do not support the ideal geometrical frustration picture. Then we have to focus on other possibility.

The inhomogeneity would play an important role for absence of magnetic ordering as disorder induces a spin-liquid-like state in frustrated spin systems \cite{Watanabe2014,Kawamura2019}. In general, increase of linewidth in strongly-correlated electrons systems, for example, $\lambda$-(ET)$_2$GaCl$_4$ \cite{Saito2018} and $\beta'$-(ET)$_2$ICl$_2$ \cite{Eto2010} salts, are not striking except near transition temperature because an electron localizes on a dimer. The inhomogeneity in $\lambda$-(STF)$_2$GaCl$_4$ may be attributed to the charge distribution because the charge distribution gives rise to the inhomogeneity of the spin density on conducting layers. Since $\lambda$-(ET)$_2$GaCl$_4$ does not show significant line broadening above $T_{\mathrm N}$, donor molecular structure may affect the emergence of the charge distribution. There is positional disorder in the Se/S atom on the STF molecules \cite{Naito1997,Mori2002a} because the position of Se atoms are unsymmetrical in the molecule as shown in Fig.\ref{crystal}(b). The positional disorder makes random potential, which can cause inhomogeneous electron localization. However, an exact diagonalization calculation for the triangular lattice in the presence of disorder shows decrease of $1/T_1$ with decreasing temperature which contradicts our results.

Another possibility is phase competition. Seo and Fukuyama suggested the $\lambda$-type salts locate near the AF phase and the spin gap phase \cite{Seo1997} considering the quantum Monte Carlo simulations for dimerized Heisenberg chains \cite{Katoh1994,Katoh1995}. From our results, increase of $1/T_1$ indicates the AF phase is very close, whereas saturating behavior of $1/T_1$ at low temperature implies gapless excitation, meaning the end point of the spin gap phase. These results suggests competing magnetic state  between the AF and the spin gap phases. We speculate 3.5 K is characteristic temperature in $\lambda$-(STF)$_2$GaCl$_4$ that quantum effect dominates because quantum effect is enhanced as characteristic length scale $\xi$ diverges by a power law ($\xi \propto T^{-1}$) in a quantum critical regime \cite{Sachdev2011a}. Saturating behavior of $1/T_1$ at low temperature is also observed in other compounds that locate near a QCP such as the Heisenberg ladder Cu$_2$(C$_5$H$_{12}$N$_2$)$_2$Cl$_4$ \cite{Chaboussant1998} and the transverse field Ising chain CoNb$_2$O$_6 $ \cite{Kinross2014}. Qualitative similarity to those materials and distinct difference from spin liquids implies quantum criticality plays a role for the suppression of the long-range ordering. To explore further properties near the quantum critical regime, experiments of $\lambda$-(ET)$_2$GaCl$_4$ under pressure is interesting to how the magnetic behavior changes with approaching the chemical pressure of $\lambda$-(STF)$_2$GaCl$_4$. Besides, theoretical studies based on this stripe lattice system are required to understand the whole aspect of phase diagrams of the stripe lattice system.

\section{CONCLUSION}
We performed site-selective $^{13}$C NMR to investigate a non-magnetic insulating phase in the stripe lattice system consisting of triangular and square tilings using $\lambda$-(STF)$_2$GaCl$_4$. $1/T_1T$ increases with decreasing temperature, indicating AF spin fluctuations develop. The origin of AF spin fluctuations can be understood by dimer-Mott insulating picture as well as the antiferromagnet $\lambda$-(ET)$_2$GaCl$_4$. These results demonstrated that AF spin fluctuations exist in the insulating phase near the SC phase of the $\lambda$ salts, which can contribute the SC pairing mechanism. Regardless of the large enhancement $1/T_1$ below 10 K, $1/T_1$ shows saturating behavior below 3.5 K and no indication of long-range magnetic ordering down to 1.63 K, which is two orders of magnitude less than the exchange interaction $J\simeq 194$ K. This is consistent with absence of NMR peak splitting and magnetic susceptibility. NMR linewidth increases with decreasing temperature, indicating the electronic state is inhomogeneous. These results suggest that $\lambda$-(STF)$_2$GaCl$_4$ is a disordered electronic system and a novel quantum disordered state is realized. We consider geometrical frustration, disorder, and the QCP between the AF phase and the spin gap phase pictures to describe non-magnetic ground state. This, to our knowledge, first observation of anomalous magnetic behavior near the AF phase in organic conductors opens new perspectives: given the similarities to the QCP rather than other quantum states, further experiments in $\lambda$-(ET)$_2$GaCl$_4$ under pressure toward the chemical pressure of $\lambda$-(STF)$_2$GaCl$_4$ is clearly called for. Besides, theoretical studies in the stripe lattice system will also help understand the quantum state.

\begin{acknowledgments}
This work was supported by JSPS KAKENHI Grant Number JP16K0542706 and the Deutsche Forschungsgemeinschaft (DFG) via grant DR228/39-3. We thank Y. Takahashi and T. Inabe at Hokkaido University for X-ray diffraction to determine crystal axes, G. G. Lesseux for ESR measurements, and G. Untereiner for preparation of resistivity measurements at University of Stuttgart.
\end{acknowledgments}

\appendix
\section{Chemical shift tensor and hyperfine coupling constant}
In general, the NMR shift $\delta$ is written as $\delta = K + \sigma = A \chi_{\mathrm s} + \sigma$, where $K$ is the Knight shift, $A$ is the hyperfine coupling constant, $\chi_{\mathrm s}$ is the spin susceptibility, and $\sigma$ is the chemical shift. The chemical shift tensor for $\lambda$ salts are calculated in the same manner as Ref.\onlinecite{Saito2015} using the crystal structure of the $\lambda$ salt \cite{Tanaka1999} and the chemical shift tensor of (ET)$^{+0.5}$ \cite{Kawai2009}. Here, we assumed the chemical shift of analog donor molecules is almost the same because the chemical shift depends mainly on the molecular structure. The resulting chemical tensors for two independent molecules are as follows:
\begin{eqnarray}
\overleftrightarrow{\sigma_{\mathrm A}}&=& \left(
\begin{array}{ccc}
\sigma_{a^*a^*} & \sigma_{a^*b} & \sigma_{a^*c}  \\
\sigma_{a^*b'} & \sigma_{b'b'} & \sigma_{b'c}  \\
\sigma_{a^*c} & \sigma_{b'c} & \sigma_{cc} 
\end{array}
\right) \\
&=& \left(
\begin{array}{ccc}
89.5 & 29.9 & 4.0 \\
29.9 & 88.3 & -23.9 \\
4.0  & -23.9 & 168.3
\end{array}
\right) ({\mathrm {ppm}}) \, ,\\
\overleftrightarrow{\sigma_{\mathrm B}}&=& \left(
\begin{array}{ccc}
89.7 & 30.4 & 2.7 \\
30.4 & 89.0 & -24.9 \\
2.7  & -24.9 & 167.5
\end{array}
\right) ({\mathrm {ppm}})\, ,
\end{eqnarray}
where $b'$ is defined as $c \times a^*$. $\sigma$ for $\theta =$45$^{\circ}$ and 0$^{\circ}$ ([110]) is estimated to be 74.6 ppm and 115.4 ppm, respectively. Here the central C=C bond length of molecule A is shorter than that of B.

\begin{figure}[tbp]
  \includegraphics[width=8.6cm]{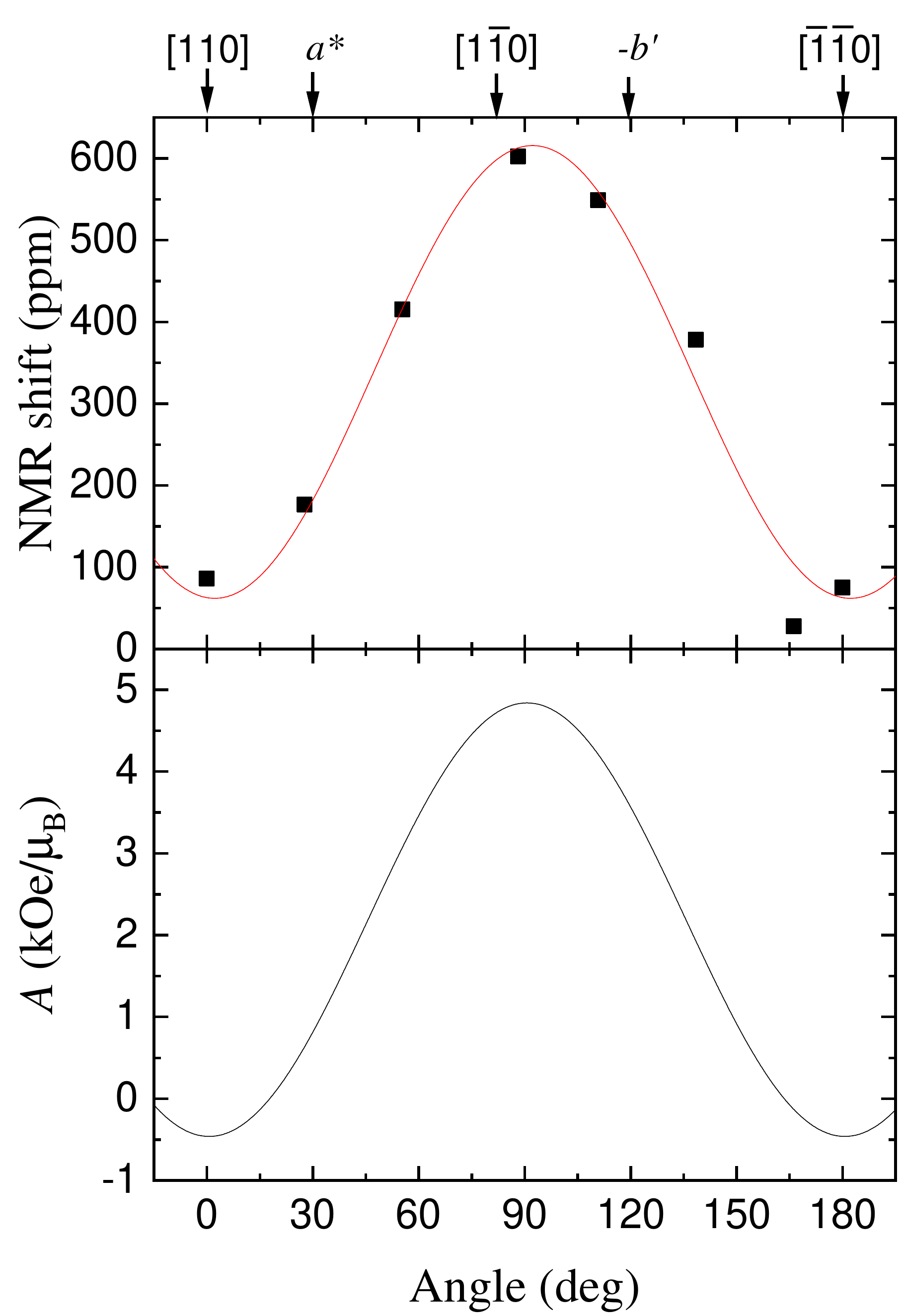}
  \caption{Angle dependence of NMR shift (upper panel) and hyperfine coupling constant (lower panel) at 100 K under $\mu_0H$ = 6 T.}\label{angle}
\end{figure}

Figure \ref{angle} shows angle dependence of the NMR shift and the hyperfine coupling constant obtanined from the NMR shift at 100 K where we used the averaged chemical shift tensor of the two molecules. We obtained hyperfine coupling constants for $\theta =$45$^{\circ}$ and 0$^{\circ}$ as $A_{45^{\circ}} = 2.1$ kOe/$\mu_\text{B}$ and $A_{0^{\circ}}= 0.46$ kOe/$\mu_\text{B}$, respectively.

By looking $\lambda$ salts' crystal shape, it is hard to distinguish $a^*$ and $b'$ axes. [1$\overline{1}0$] and [110] directions that correspond to perpendicular and parallel to the wide plane of the crystal shape is easier to distinguish in $\lambda$-STF and BETS salts. Noted that the directions are opposite against the crystal shape in $\lambda$-(ET)$_2$GaCl$_4$.

\section{DC resistivity}
\begin{figure}[htbp]
  \includegraphics[width=8.6cm]{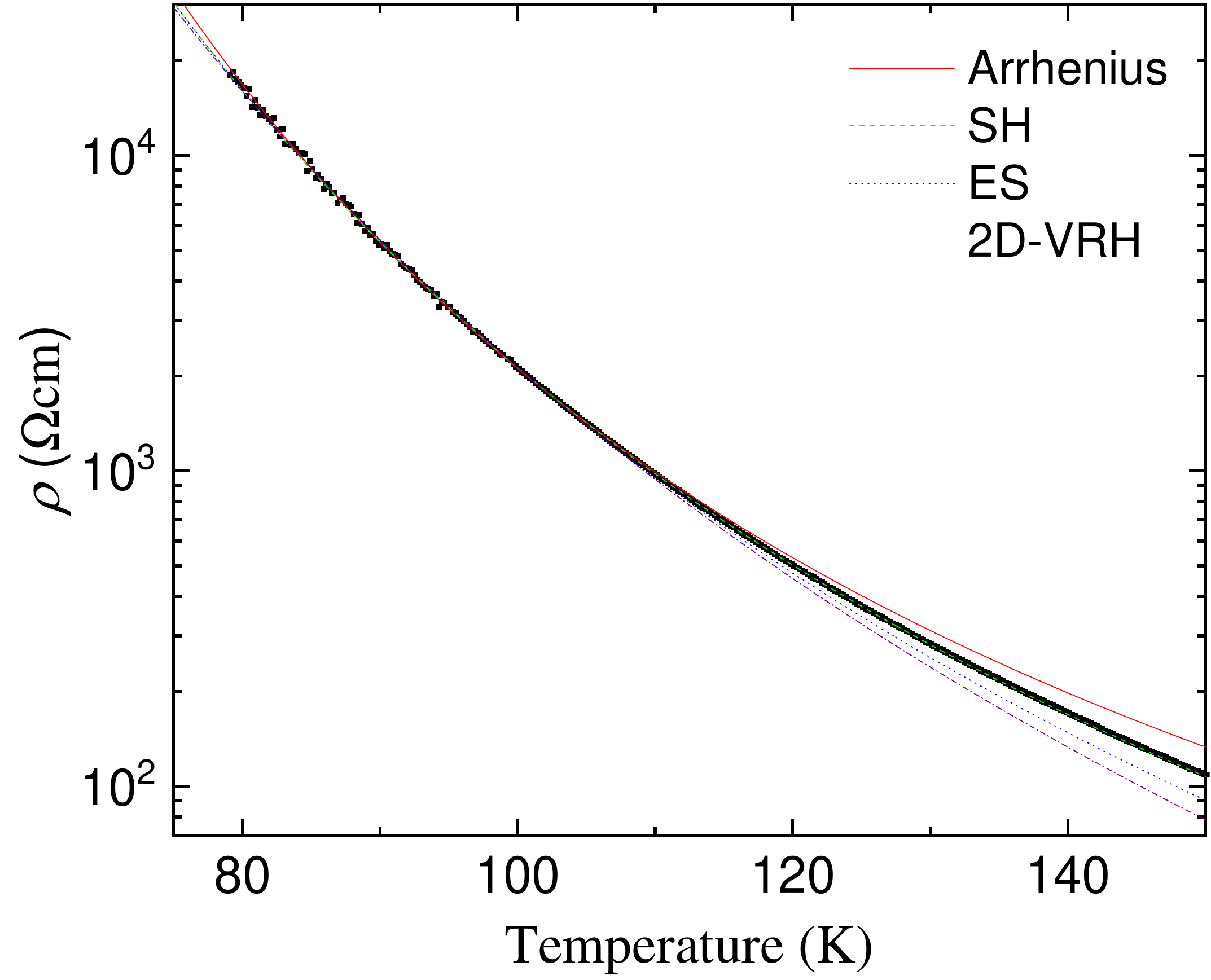}
  \caption{Temperature dependence of resistivity for the $c$ axis of $\lambda$-(STF)$_2$GaCl$_4$ with fitting curves for Arrhenius, SH, ES, and 2D-VRH models.}\label{conductivity}
\end{figure}

To characterize $\lambda$-(STF)$_2$GaCl$_4$, DC resistivity measurements were performed for the $c$ axis. The results coincide with the previous study \cite{Mori2001}. We analyzed temperature dependence with localized models as shown in Fig. \ref{conductivity}. Fit was performed in 85 K -- 125 K using the following equations: the Arrhenius law, the variable-range-hopping model for two dimension, and the Efros and Shklovskii (ES) model, and the soft Hubbard (SH) gap model where the SH model takes short-range Coulomb interaction into account proposed by Shinaoka and Imada \cite{Shinaoka2010}, where models except the Arrhenius law are usually applied in disordered systems. Below 100 K, all models fit well with experimental results. However, Arrhenius, ES, and 2D-VRH models deviate above 100 K whereas the SH model covers wide temperature range even though high temperature is not included in the fitting range. The better fit result of the SH model suggests that the electronic state is inhomogeneous, which is consistent with NMR results.

%

\end{document}